\documentclass[onecollarge,natbib]{svjour2}
\bibpunct{[}{]}{;}{n}{}{,} 
\smartqed  
\usepackage{graphicx}
\usepackage{enumerate}
\usepackage{multirow}
\usepackage{longtable}
\usepackage{epsfig}
\usepackage[latin1]{inputenc}
\usepackage[english]{babel}
\usepackage{array}
\usepackage{subfigure}
\usepackage{mathptmx}

\journalname{Few-Body Systems}
\begin{document}

\title{A theoretical analysis of mid-mass neutron halos}
\subtitle{What changes going from few- to many-body systems regarding neutron halos?}


\author{T. Duguet     
}


\institute{T. Duguet \at
              Centre de Saclay, IRFU/Service de Physique Nucl\'eaire, F-91191 Gif-sur-Yvette, France and\\
KU Leuven, Instituut voor Kern- en Stralingsfysica, 3001 Leuven, Belgium and\\
National Superconducting Cyclotron Laboratory and Department of Physics and Astronomy,
Michigan State University, East Lansing, MI 48824, USA
              \email{thomas.duguet@cea.fr}           
}

\date{Received: date / Accepted: date}

\maketitle

\begin{abstract}
The present contribution summarizes the content and slightly updates the discussion of a recently proposed theoretical analysis of the halo phenomenon in many-fermion systems. We focus here on applications to potential neutron halos in mid-mass nuclei. 
\end{abstract}

\section{Introduction}
\label{intro}

The study of light halo nuclei at the limit of stability, i.e. light weakly-bound systems in which either the proton or the neutron density displays an
unusually extended tail~\cite{hansen87}, has been possible in the last three decades thanks to first generations of radioactive ion beam facilities. Since the first experimental observation of this exotic structure in \mbox{$^{11}$Li}~\cite{tanihata85a,tanihata85b}, several light neutron halo systems have been identified, e.g. \mbox{$^6$He}~\cite{zhukov93},
\mbox{$^{11}$Be}~\cite{tanihata88,fukuda91,zahar93}, \mbox{$^{14}$Be}~\cite{tanihata88,thompson96},
\mbox{$^{17}$B}~\cite{tanihata88} or \mbox{$^{19}$C}~\cite{bazin95,kanungo00}. 

As the neutron drip-line is only known experimentally up to oxygen~\cite{Thoennessen:2013pna,Thoennessen:2014hia,Thoennessen:2015hfa}, it is a compelling question whether halo systems can exist in heavier nuclei. Indeed, one may wonder whether the mass increase and the associated collectivity will hinder the occurrence of halos? If so, what would be the corresponding typical mass limit beyond which one would observe nuclei displaying a less exotic neutron skin rather than a genuine halo structure? Eventually, if halos were to emerge in systems heavier than p-shell or light sd-shell nuclei, would they display different structural properties, e.g. would the standard cluster picture at play in light halo nuclei be pertinent to describe mid-mass neutron halo nuclei? 

These are some of the questions one may raise today regarding potential mid-mass halo nuclei. As mentioned above, it is only possible to speculate theoretically at this point in time given that experimental answers will only start to arise with the upcoming generation of radioactive ion beam facilities. To do so, one needs a robust structure analysis tool that is as model-independent as possible and that applies to any type of many-fermion systems. The first aim of the present contribution is to summarize one such analysis method proposed recently~\cite{Rotival:2007hp}. The method in question (i) is solely based on the one-nucleon density distribution and (ii) applies to all systems from non halos to paradigmatic halos. The method eventually translates into quantitative halo parameters that lead to identifying three typical energy scales for paradigmatic halo systems and that allow one to disentangle between neutron skins and genuine halos. The second aim of the present contribution is to summarize and slightly update the discussion of the results presented in Refs.~\cite{Rotival:2007sy} that shed light on some of the questions raised above, although much remains to be done for a clear picture to emerge regarding halo systems beyond the lightest nuclei.

\section{Analysis method}
\label{analysis}

A useful method to study halos in many-body systems be capable of characterizing, in a model-independent fashion, a component of the nucleon density distribution that is spatially decorrelated from an a priori unknown core. As summarized below, it is indeed possible to propose quantitative criteria to identify and characterize halos starting from an analysis of medium- and long-distance properties of the one-nucleon (internal) density.

\subsection{Mid- and long-range properties of the one-nucleon density distribution}
\label{decompositiondensity}

Generally speaking, the electromagnetic or weak charge density (and current) operators are expressed as
an expansion in many-body operators acting on nucleonic degrees of freedom, i.e. it reads in first quantization form as
\begin{eqnarray}
\label{eq:fourvec}
\hat{\rho}_{{\rm ch}}(\vec{q})  &=& \sum_{i=1}^{N} \hat{\rho}_{{\rm ch}}(i;\vec{q}) + \frac{1}{2} \sum_{i,j=1}^{N} \hat{\rho}_{{\rm ch}}(i,j;\vec{q}) + \dots\ , 
\end{eqnarray}
where $\vec{q}$ is the momentum associated with the external probe. Staying qualitative for simplicity, this remark is meant to point that limiting oneself to the one-body operator and thus neglecting meson-exchange currents constitutes an approximation of a more systematic scheme~\cite{Horowitz01a}. This may eventually impact the extraction of the point, e.g. neutron, density distribution that we are presently interested in and whose associated operator is given by $\hat{\rho}(\vec{r}) \equiv \sum_{i=1}^{N} \delta(\vec{r}-\hat{\vec{r}}_i)$.

The point neutron density distribution of an even-even $J^{\pi}=0^+$ ground-state can be expanded as\footnote{The discussion can be conducted thoroughly for the {\it internal} density~\cite{Rotival:2007hp} but is presently applied to the lab-frame one-body density for simplicity. Also, the intrinsic spin is omitted from most of the presentation for simplicity as it can be easily reinserted when reaching final results and drawing conclusions.}
\begin{equation}
\rho(\vec{r})\equiv \frac{\langle \Phi^{N}_{0} |\hat{\rho}(\vec{r})| \Phi^{N}_{0}\rangle}{\langle \Phi^{N}_{0} | \Phi^{N}_{0}\rangle} = \sum_{\nu \in {\cal H}_{N-1}}\frac{2l_\nu+1}{4\pi}|\bar{\varphi}_\nu(r)|^2 \, , \label{eq:decomp_density4}
\end{equation}
where $\bar{\varphi}_\nu(r)$ denotes the radial part of the overlap function $\varphi_\nu(\vec{r}\,) \equiv \langle \Phi^{N}_{0}  | a^{\dagger}(\vec{r}\,) | \Phi^{N-1}_{\nu} \rangle$, i.e. the density distribution is spherically symmetric. The spectroscopic factor associated with state $| \Phi^{N-1}_{\nu} \rangle$ denotes the norm of the corresponding overlap function
\begin{equation}
S_\nu=\int d\vec{r}\,|\varphi_\nu(\vec{r}\,)|^2\, ,
\end{equation}
such that, introducing normalized overlap functions $\psi_\nu(\vec{r}\,)$, Eq.~\ref{eq:decomp_density4} is rewritten as
\begin{equation}
\rho(r)=\sum_{\nu} \frac{2l_\nu+1}{4\pi}\,S_\nu\,\left|\bar{\psi}_\nu(r)\right|^2\equiv\sum_{\nu}C_\nu(r)
\label{eq:decomp_density2}\,.
\end{equation}

At large distances effects of inter-nucleon interactions vanish such that, for $r\longrightarrow+\infty$, one has
\begin{eqnarray}
\rho(r) & \longrightarrow &\sum_{\nu \in {\cal H}_{N-1}}\frac{B_{\nu}^2}{4\pi}(2l_\nu+1)|h_{l_\nu}(i\,\kappa_\nu
r)|^2 \label{eq:rho_asympt_usual1} \\ 
& \longrightarrow &\sum_{\nu \in {\cal H}_{N-1}}\frac{B_{\nu}^2}{4\pi}(2l_\nu+1) \frac{e^{-2\kappa_\nu r}}{(\kappa_\nu r)^2} \label{eq:rho_asympt_usual2} \\ 
& \longrightarrow &\frac{B_{0}^2}{4\pi}(2l_0+1)\frac{e^{-2\kappa_0\,r}}{(\kappa_0\,r)^2} \,, \label{eq:rho_asympt_usual3} 
\end{eqnarray}
with $\kappa_\nu\equiv \sqrt{-2mE^{-}_\nu/\hbar^2}$, where $E^{-}_\nu\equiv\left(E_0^{N}-E_\nu^{N-1}\right)$ denotes minus the one-neutron separation energy to reach $| \Phi_\nu^{N-1} \rangle$, $h_{l_\nu}(i\,\kappa_\nu r)$ the Hankel function\footnote{The asymptotic of proton and neutron densities (Hankel functions for neutrons, Whittaker functions for protons) are different because of the charge factor. We presently focus on neutron halos and thus on the neutron density.} and $B_{\nu}$ the asymptotic normalization coefficient (ANC). Note that the more bound the overlap function, the more excited the corresponding state in the $(N-1)$-body system.

\subsection{Crossing pattern}
\label{sec:newcrit_asympt_cons}

To leading order, overlap functions are ordered at very large distances according to their separation energies $|E^{-}_\nu|$. Corrections to this ordering at smaller distances come from (i) the $l_{\nu}$-dependence of the Hankel functions due to the centrifugal
barrier, which favors low angular-momentum states, and (ii) the $(2l_{\nu}+1)$ degeneracy factor which favors
high angular-momentum states. In any case, for extremely large distances the least bound component will always prevail; see Ref.~\cite{Rotival:2007hp} for a detailed discussion. This long-distance ordering has interesting consequences on the properties of the density as a whole as it induces a typical crossing pattern at shorter distances as is now briefly explained.

Let us first omit spectroscopic factors, i.e. let us equate them to one. As Eqs.~\ref{eq:rho_asympt_usual3} and~\ref{eq:decomp_density2} testify, the \mbox{$\nu=0$} overlap function corresponding to the smallest separation energy dominates at large distances. Because of continuity and normalization conditions, it implies that \mbox{$\bar{\psi}_0(r)$} must cross the other overlap functions as $r$ goes inward from $+\infty$ to zero. The position at which $\bar{\psi}_0$ crosses each $\bar{\psi}_\nu$ depends on the difference of their separation energies and on their angular momenta. In
particular, there will exist a crossing between \mbox{$|\bar{\psi}_0(r)|^{2}$} and the remaining density
\mbox{$\left[\rho(r)-C_0(r)\right]$}. The same is true about \mbox{$|\bar{\psi}_1(r)|^{2}$}: it must cross the
remaining density \mbox{$\left[\rho(r)-C_0(r)-C_1(r)\right]$}... As a result, any given individual component
must cross the sum of those that are more bound. Of course, the centrifugal barrier influences the position of such
crossings but not their occurrence because of the robustness of the (very) asymptotic ordering pattern discussed above.

Spectroscopic factors $S_\nu$ are known to increase with the excitation energy of the corresponding eigenstate $| \Phi^{N-1}_{\nu} \rangle$. Thus, the norm of $\varphi_0$ is smaller than for more excited components $\varphi_\nu$, which mechanically ensures the existence of the
crossings discussed previously. A similar reasoning holds when going from $\varphi_0$ to $\varphi_1$ etc. See Ref.~\cite{Rotival:2007hp} for a more detailed discussion.

\begin{figure}[hptb]
\includegraphics[keepaspectratio, angle = -90, width = 0.5\columnwidth]{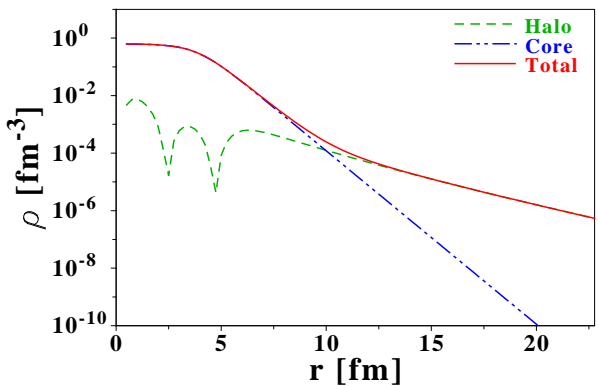}
\caption{\label{fig:model} (Color Online) "Core+tail" simplified model. The total density is the superposition of a well-bound component and a
loosely-bound one. A semi-phenomenological density is used for the core density, whereas the halo part is the realistic $3_{1/2}$ state of $^{80}$Cr obtained from a SR-EDF calculation. Taken from Ref.~\cite{Rotival:2007hp}.}
\end{figure}

\subsection{Relevant energy scales}
\label{energyscales}

A nuclear halo emerges from a substantial probability for nucleons to partially occupy a spatial region that is remote ("separate") from an a priori unknown inner ("core") part. This can only be achieved if some overlap functions contributing to the density exhibit very long tails. Most importantly, the delocalization from the core requires the latter to exist and to remain well localized. To achieve this, it is necessary to have a crossing between two well-identified groups of overlap functions with significantly different asymptotic slopes. This crossing between two groups of orbitals translates into a curvature in the density. The sharper the crossing, the larger the curvature and the more pronounced the halo. The situation is schematically illustrated in Fig.~\ref{fig:model} for a simple model where the halo is due to a
single overlap function. Of course, more realistic situations must be considered in connection with the greater collective character of medium-mass systems implying that makes hardly probable for a single overlap function to be well separated from all the others.

The need for the existence of two groups of orbitals characterized by significantly different asymptotic slopes provides critical conditions for the appearance of a halo: (i) the least bound component $\varphi_0$ must have a very small separation energy to extend far out, (ii)
several components $\varphi_1, \varphi_2\ldots \varphi_m$ may contribute significantly to the density tail
if, and only if, they all have separation energies of the same order as that of $\varphi_0$, (iii) for this tail to
be spatially {decorrelated} from the rest of the density (the "core"), the components with $\nu>\nu_m$ have
to be much more localized than those with $\nu\le \nu_m$. This third condition is fulfilled when the
crossing between the $m^{th}$ and $(m+1)^{th}$ components in the density is sharp, which corresponds to
significantly different decay constants $\kappa_m\ll\kappa_{m+1}$ at the crossing point.

The ideal situation just discussed translates into specific patterns in the excitation energy spectrum of the $(N-1)$-body
system. It suggests that a halo appears when (i) the one-neutron separation energy
$S_n=|E^{-}_0|$ is close to zero, (ii) a bunch of low-energy states have separation energies $|E^{-}_\nu|$ close to zero as well, and (iii) a significant gap in the spectrum of the $(N-1)$-body system exists, which separates the latter bunch of states from higher
excitations.

\begin{figure}[hptb]
\includegraphics[keepaspectratio,width=0.5\columnwidth]{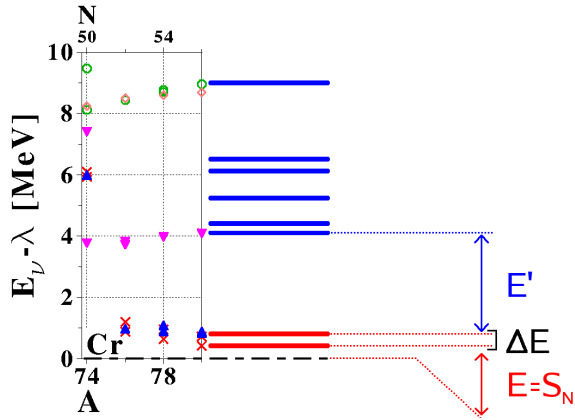}
\caption{\label{fig:EFT_spectrum} (Color Online) Separation energy spectrum $|E^{-}_\nu|$
 of the $(N-1)$-body system relative to the N-body ground state. (Right) Schematic display of the energy scales relevant for the appearance of a halo. (Left) Realistic spectrum obtained from  SR-EDF calculations for the four last bound chromium isotopes (see Sec.~\ref{sec:res_cr} for more details). Taken from Ref.~\cite{Rotival:2007hp}.}
\end{figure}

A similar discussion was given in the context of designing an effective field theory (EFT) for light weakly-bound
nuclei~\cite{bertulani02}, where two energy scales $(E,E')$ were found to be relevant: (i) the nucleon separation
energy $E=S_n$ that drives the asymptotic behavior of the one-body density, and (ii) the core excitation
energy $E'=|E^{-}_{m+1}|$ that needs to fulfill $E'\gg E$ for the tail orbitals to
be well decorrelated from the remaining core. The additional energy scale that we presently identify is the energy
spread $\Delta E$ of (possible) low-lying states of the $(N-1)$-body system, which becomes relevant when more than
one component is involved in the halo. The corresponding picture is displayed in Fig.~\ref{fig:EFT_spectrum}.

More quantitatively, the ideal situation for the formation of a halo is obtained for (i) a very small separation
energy, in orders of a few hundred keVs, the empirical value of \mbox{$2$~MeV$/A^{2/3}$}~\cite{fedorov93,jensen00} giving a good approximation of expected values, (ii) a narrow bunch of low-lying
states, whose spread should not exceed about $1$~MeV, and (iii) a large gap with
higher-excited states, at least four or five times the separation energy. Those values are of course only indicative, knowing
that there is no sharp limit between halo and non-halo domains.

\subsection{Halo region}
\label{sec:newcrit_def_region}

\begin{figure}[hptb]
\includegraphics[keepaspectratio, width = 0.5\columnwidth]{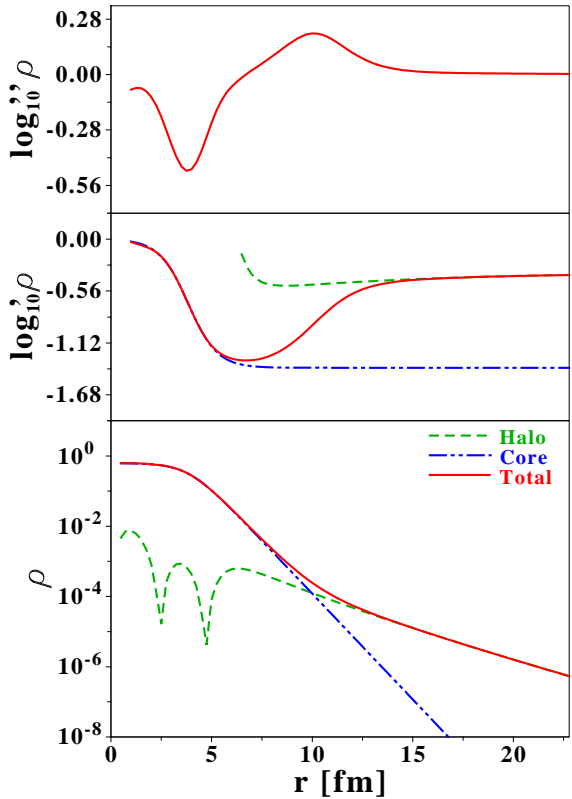}
\caption{\label{fig:crit_qualit_1} (Color Online) Ankle in the (base-$10$) log-density: log-density (bottom panel), first (middle panel) and second (top panel) log-derivatives. Taken from Ref.~\cite{Rotival:2007hp}.}
\end{figure}

The (more or less) sharp ankle in the density due to the crossing between (more or less) aggregated low-lying components and upper-lying ones translates into a (more or less) pronounced peak in the second derivative of the (base-$10$) logarithmic profile ($\log_{10}$) of the one-body density. This is illustrated in Fig.~\ref{fig:crit_qualit_1} for the schematic model introduced in Fig.~\ref{fig:model}.

At radius \mbox{$r=r_{max}$} corresponding to the maximum of that peak, core and tail contributions cross,
i.e. they contribute equally to the total density. At larger radii, the halo, if it exists, dominates. Therefore,
\textit{the spatially decorrelated region is defined as the region beyond the radius $r_0$ where the core density
is one order of magnitude smaller than the halo one}. This definition raises immediately two questions. First, the choice of one order of magnitude is arbitrary. However, its relevance as a conservative choice will appear later on when introducing quantitative halo criteria and computing them for sequences of nuclear systems. Second, one only extracts/computes the (neutron) density as a whole in practice such that $r_0$ cannot be accessed directly. Extensive simulations have been performed in Ref.~\cite{Rotival:2007hp} to address this key point. Given $r_{max}$, which can be extracted from the total neutron density, $r_0$ was found to be reliably specified through
\begin{equation}
\frac{\partial^2 \log_{10}{\rho(r)}}{\partial r^2}|_{r=r_0}\equiv\frac{2}{5}\,\frac{\partial^2 \log_{10}{\rho(r)}}{\partial r^2}|_{r=r_{max}} \hspace{0.6cm} {\rm with} \hspace{0.6cm} r_0>r_{max} \, .  \label{eq:def_r0}
\end{equation}
Once validated, the method to isolate the halo region ($r\geq r_0$) only relies on the total density as an input, and does not require an \textit{a priori}
separation of the one-body density into core and halo parts. See Ref.~\cite{Rotival:2007hp} for details, where a tolerance on the estimation of $r_0$ is further discussed.

\subsection{Quantitative halo parameters}
\label{haloparam}

With $r_0$ at hand, several quantitative measures are introduced to characterize the presence (absence) of a neutron halo. First, the average number of
neutrons in the halo region can be extracted through\footnote{For neutron-rich medium-mass nuclei, protons are well confined in the nuclear interior and do not participate in the long-range part of the total density $\rho$. The total density or the neutron density can thus equally be used to evaluate $N_{\mathrm{halo}}$ and $\delta R_{\mathrm{halo}}$. See Ref.~\cite{Rotival:2007hp}.}
\begin{equation}
\label{eq:def_nhalo} N_{\mathrm{halo}}\equiv 4\pi\int_{r_0}^{+\infty}\!\!\rho(r)\,r^2\,dr \, \, \, .
\end{equation}
Second, the impact of the halo on the matter or neutron r.m.s. radius of the nucleus can be evaluated through\footnote{Note that the tolerance margins on $r_0$ propagates into theoretical uncertainties on $N_{\mathrm{halo}}$ and $\delta R_{\mathrm{halo}}$.}
\begin{eqnarray}
\delta R_{\mathrm{halo}}&\equiv&R_{\mathrm{r.m.s.},tot}-R_{\mathrm{r.m.s.},inner}=\!\!\!\!\sqrt{\frac{\int_{0}^{+\infty}\rho(r)r^4\,dr}
{\int_{0}^{+\infty}\rho(r)r^2\,dr}}-
\sqrt{\frac{\int_{0}^{r_0}\rho(r)r^4\,dr}
{\int_{0}^{r_0}\rho(r)r^2\,dr}}\,,
\label{eq:def_drhalo}
\end{eqnarray}
knowing that extensions to	any radial moment of the density can be envisioned\footnote{Numerical issues appear when going to high-order moments. Indeed,
\mbox{$\langle r^n \rangle$} is more and more sensitive to the upper limit of integration as $n$ increases. Thus,
the result may significantly depend on the box size used to discretize the continuum or on the size of the basis
used to expand the many-body solution.}. Quantities $N_{\mathrm{halo}}$ and $\delta R_{\mathrm{halo}}$ are of course correlated, but they do not carry exactly the same information as will be illustrated later on.

Third, a more detailed characterization of the halo can be achieved by computing the contributions of each overlap function to it via
\begin{equation}
N_{\mathrm{halo},\nu}\equiv4\pi \,
(2j_\nu+1)\,\int_{r_0}^{+\infty}|\bar{\varphi}_{\nu}(r)|^2\,r^2\,dr\,.\label{eq:def_nhalo_i}
\end{equation}

\section{Applications to mid-mass neutron-rich nuclei}
\label{appli}

The analysis method introduced in Sec.~\ref{analysis} is versatile as it applies to many-fermion systems characterized by different constituents and scales, as long as they are governed by finite-range interactions. In Ref.~\cite{Rotival:2007sy}, the method was successfully applied to light nuclei studied through coupled-channels calculations~\cite{nunes96a,nunes96b}, to medium-mass nuclei described through single-reference energy density functional (SR-EDF) calculations~\cite{bender03b,Duguet:2013dga} and to atom-positron/ion-positronium complexes computed through the fixed-core stochastic variational method~\cite{varga95,ryzhikh98b,mitroy01}. In the present contribution, we focus on mid-mass neutron-rich nuclei computed via SR-EDF calculations.

The nuclear EDF approach is the microscopic tool of choice to study medium-mass and heavy nuclei in a systematic manner~\cite{bender03b} and it is presently considered in its single-reference implementation~\cite{Duguet:2013dga}.  Calculations are performed using the non-relativistic code HFBRAD~\cite{bennaceur05a} that solves Hartree-Fock-Bogoliubov equations in  spherical symmetry, discretizing coordinate space  within a sphere using vanishing boundary conditions for the wave functions (Dirichlet conditions). Convergence of the calculations as a function of numerical parameters
has been checked for all results presented here. Except if stated otherwise, calculations employ a Skyrme SLy4 functional~\cite{chabanat97,chabanat98} in the particle-hole channel and a density-dependent delta interaction (DDDI) corresponding to a "mixed-type" pairing (see Sec.~\ref{sec:pair_loc}) to generate a purely local functional in the particle-particle channel. See Refs.~\cite{Rotival:2007hp,Rotival:2007sy} for details.

In the present section, the analysis method is illustrated on drip-line chromium (tin) isotopes as paradigmatic neutron (non) halo nuclei.

\subsection{Cr isotopes}
\label{sec:res_cr}

\begin{figure}[hptb]
\includegraphics[keepaspectratio,angle = -90, width = 0.5\columnwidth]{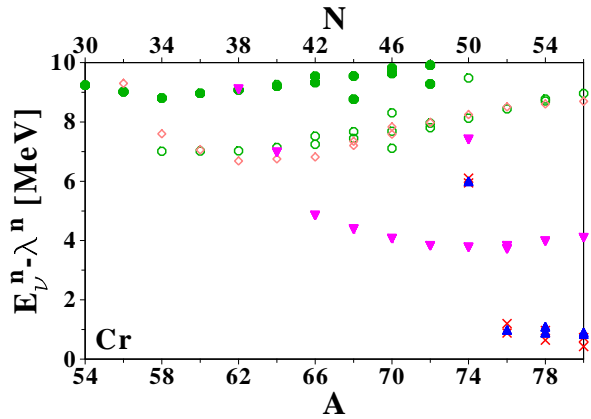}
\caption{ \label{fig:Cr_hfb_spectrum} (Color Online) Neutron separation energies \mbox{$|E^{-}_\nu|$} along the
Cr isotopic chain for final states with spectroscopic factors \mbox{$S^n_\nu>0.01$}. Conventions to label final $\nu=J^{\pi}$ states of odd-even isotopes are given in Tab.~\ref{conventions}. Taken from Ref.~\cite{Rotival:2007hp}.}
\end{figure}

The one-neutron separation energy spectrum \mbox{$|E^{-}_\nu|$} between the $J^\pi=0^+$ N-body ground state of even-even chromium isotopes and eigenstates of the \mbox{$(N-1)$-body} isotopes is shown in Fig.~\ref{fig:Cr_hfb_spectrum}. Table~\ref{tab:Cr_spect} details this information for $^{79-80}$Cr. The separation energy to the $J^{\pi}=1/2^{+}$ ground state of $^{79}$Cr is \mbox{$S_n=|E^{-}_0|\approx430$}~keV, whereas four excited states ($J^{\pi}=1/2^{+}$ and $J^{\pi}=5/2^{+}$) are within an energy spread \mbox{$\Delta E\approx470$}~keV, and are further separated from higher-excited states by \mbox{$E'\approx3.2$}~MeV. The one-neutron separation energy $S_n$ of \mbox{$^{80}$Cr} is thus compatible with the
phenomenological estimates necessary for the appearance of light halo nuclei, namely \mbox{$2$~MeV$/A^{2/3}\approx137$~keV} for $A=80$.
According to the discussion of Sec.~\ref{energyscales}, the energy scales at play in the three last bound Cr isotopes correspond to ideal halo candidates.

\begin{table}
\begin{tabular}{|l|c||l|c|}
\hline
Crosses & $\nu=1/2^{+}$ & & \\
\hline
Empty circles & $\nu=1/2^{-}$ & Full circles & $\nu=3/2^{-}$ \\
\hline
Empty upper triangles & $\nu=3/2^{+}$ & Full upper triangles & $\nu=5/2^{+}$ \\
\hline
Empty diamonds & $\nu=5/2^{-}$ & Full diamonds & $\nu=7/2^{-}$ \\
\hline
Empty lower triangles & $\nu=7/2^{+}$ & Full lower triangles & $\nu=9/2^{+}$ \\
\hline
Empty semi circles & $\nu=9/2^{-}$ & Full semi circles & $\nu=11/2^{-}$ \\
\hline
Empty squares & $\nu=11/2^{+}$ & Full squares & $\nu=13/2^{+}$ \\
\hline
\end{tabular}
\label{conventions}
\caption{Conventions used in all the figures to label final $\nu=J^{\pi}$ states of odd-even isotopes.}
\end{table}

\begin{table}
\begin{tabular}{rll}
\hline 
&&\\
& $J^{\pi}$ & $|E^{-}_\nu|$~[MeV] \\
&&\\
\hline
&&\\
&   & $>10$ \\
& $1/2^{-}$ & 9.0 \\
& $5/2^{-}$ & 8.7 \\
& $9/2^{+}$ & 4.1 \\
\multirow{2}{*}{$E'\left\updownarrow
\vphantom{\begin{array}{l}a\\a\end{array}}\right.$}  &\\
&& \\
&& \\
\multirow{4}{*}{$\Delta E\left\{\vphantom{\begin{array}{l}a\\a\\a\\a\end{array}}\right.$} 
&$5/2^{+}$ & 0.9 \\
&$5/2^{+}$ & 0.8  \\
&$1/2^{+}$ & 0.7 \\
&$1/2^{+}$ & 0.4 \\
$E\updownarrow$& & \\
 \hline
\end{tabular}
\caption{\label{tab:Cr_spect} (Color Online) One-neutron separation energies \mbox{$|E^{-}_\nu|$} from the ground state of $^{80}$Cr to final states of $^{79}$Cr with spectroscopic factors greater than $10^{-2}$.}
\end{table}

\begin{figure}[hptb]
\includegraphics[keepaspectratio,angle = -90, width = 0.5\columnwidth]{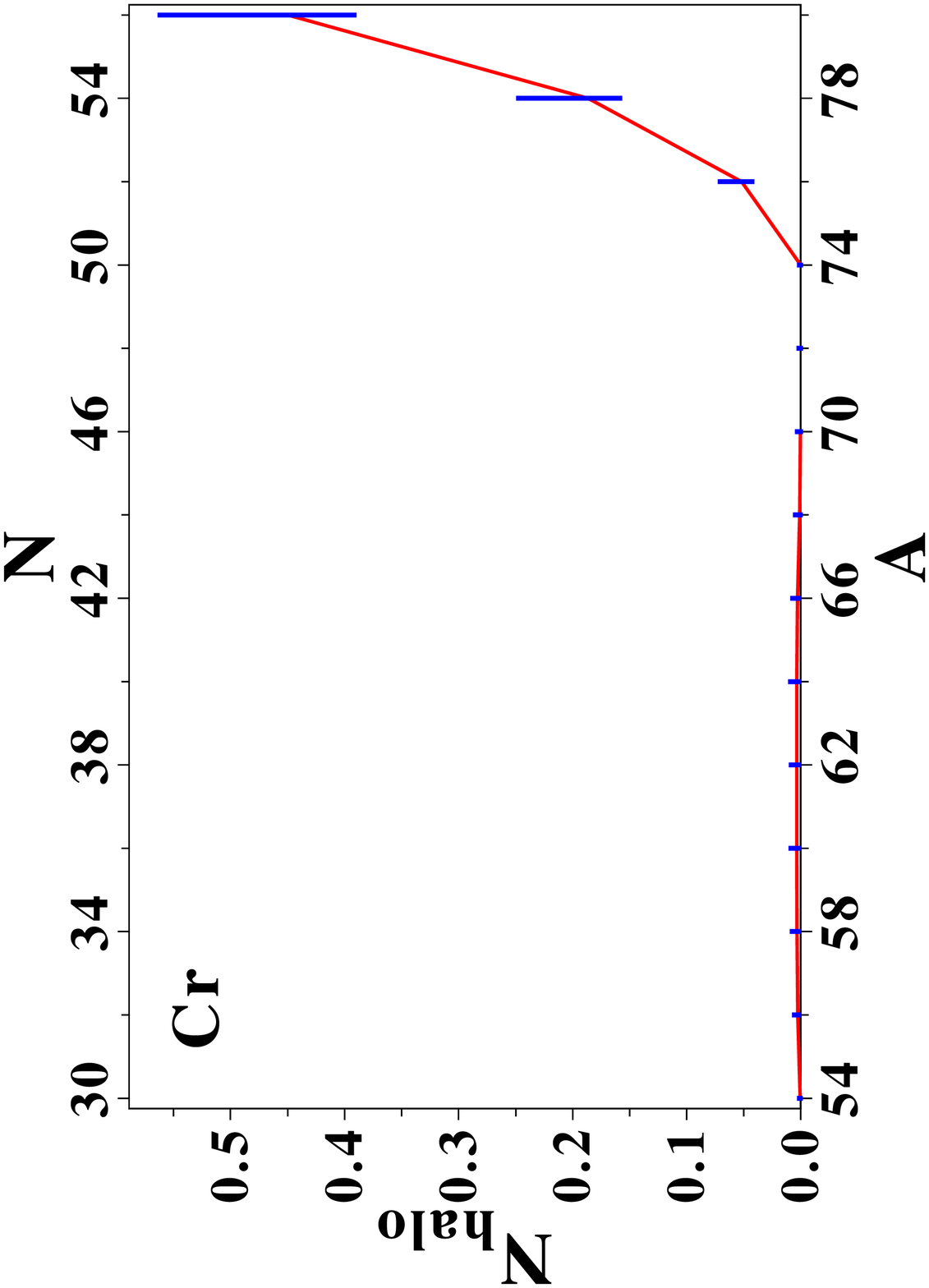}
\caption{(Color Online) Average number of nucleons participating in the halo along
the Cr isotopic chain, as a function of the nuclear mass, as
predicted by the \{SLy4+REG-M\} functional.  Taken from Ref.~\cite{Rotival:2007hp}.} \label{fig:Cr_Nhalo}
\end{figure}

\begin{figure}[hptb]
\includegraphics[keepaspectratio,angle = -90, width = 0.5\columnwidth]{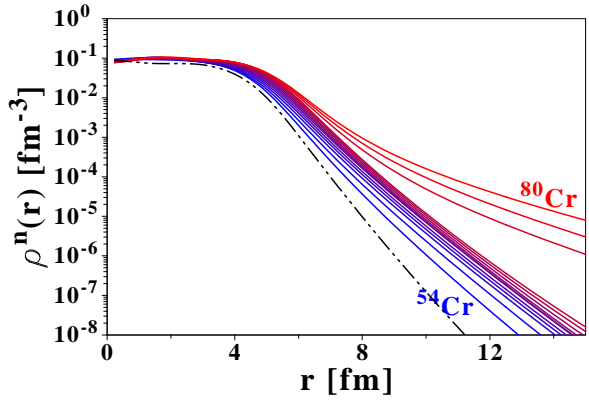}
\caption{(Color Online) Neutron densities for even-even Cr isotopes, from \mbox{$^{54}$Cr}
to \mbox{$^{80}$Cr}. The proton density of \mbox{$^{54}$Cr} is given (dashed-dotted line) as a reference for the neutron skin. Taken from Ref.~\cite{Rotival:2007hp}.} \label{fig:Cr_alldens}
\end{figure}

The quantitative measures introduced in Sec.~\ref{haloparam} can now be computed. Figure~\ref{fig:Cr_Nhalo} shows the average number of nucleons participating in the potential halo region. Whereas $N_{\mathrm{halo}}$ is consistent with zero for \mbox{$N\le50$}, a sudden increase is seen beyond the \mbox{$N=50$} shell closure. The existence of a decorrelated region in the density of the last three Cr isotopes is consistent with the
evolution of the neutron densities along the isotopic chain displayed in Fig.~\ref{fig:Cr_alldens}. The value of $N_{\mathrm{halo}}$ remains small in comparison to the total neutron number, as the decorrelated region is populated by \mbox{$\sim0.45$} nucleons on the average in \mbox{$^{80}$Cr}. In absolute value however, $N_{\mathrm{halo}}$ is comparable to what is found in light \mbox{$s$-wave} halo nuclei like \mbox{$^{11}$Be}, where roughly $0.3$ nucleons constitute the decorrelated part of the density~\cite{nunespriv}.

\begin{figure}[hptb]
\includegraphics[keepaspectratio,angle = -90, width = 0.5\columnwidth]{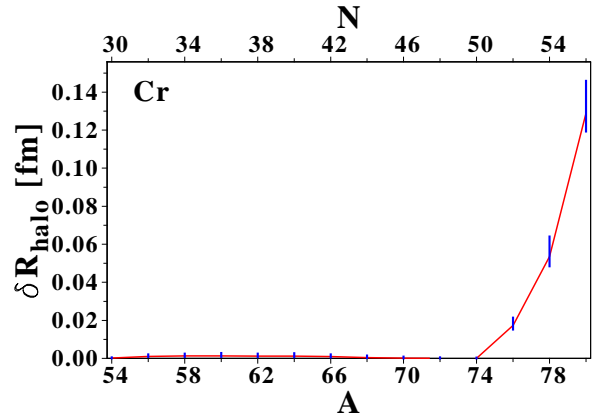}
\caption{(Color Online) Halo factor parameter $\delta{R}_{\mathrm{halo}}$ in the Cr isotopic
chain. Taken from Ref.~\cite{Rotival:2007hp}.} \label{fig:Cr_deltaR}
\end{figure}

\begin{figure}[hptb]
\includegraphics[keepaspectratio,angle = -90, width = 0.5\columnwidth]{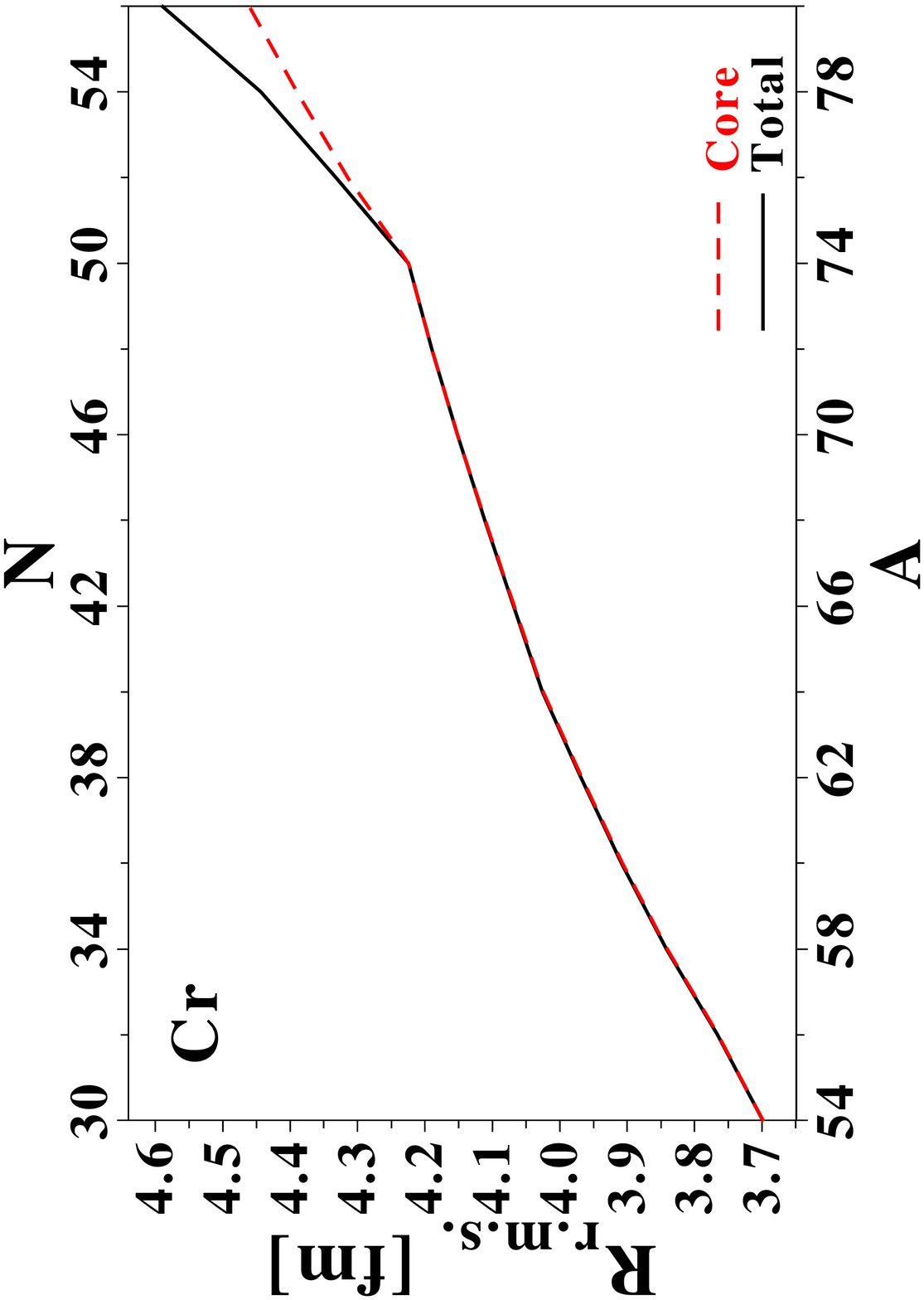}
\caption{(Color Online) Total neutron root-mean-square radius (solid line) and core contribution
(dashed line) for chromium isotopes, as predicted by the
\{SLy4+REG-M\} functional. Taken from Ref.~\cite{Rotival:2007hp}.} \label{fig:Cr_helmlike}
\end{figure}

The halo parameter $\delta{R}_{\mathrm{halo}}$ is shown in Fig.~\ref{fig:Cr_deltaR}. The halo contributes
significantly to the total neutron r.m.s. radius (up to $\sim0.13$~fm) beyond the \mbox{$N=50$} shell closure.
The latter result can be recast as a splitting of the total r.m.s. radius into a core and a halo contributions, as
displayed in Fig.~\ref{fig:Cr_helmlike}. Shell effects are properly
separated from halo ones. Indeed, the core r.m.s. radius includes a kink at \mbox{$N=50$} that is correlated to a jump in the two-neutron separation energy (not shown here) and accounts for the development of a neutron skin~\cite{Rotival:2007hp}. Contrarily, only the physics related to the existence of truly
decorrelated neutrons is extracted by $N_{\mathrm{halo}}$ and $\delta R_{\mathrm{halo}}$. 

To further characterize the halo region, individual contributions $N_{\mathrm{halo,\nu}}$ can be evaluated. As expected, the main contributions to the halo come from the lowest energy states bunched within the energy spread $\Delta E$ (if any), while for non-halo nuclei, like \mbox{$^{74}$Cr}, all contributions are consistent with zero. At the neutron drip-line, important contributions are found from both $1/2^{+}$ and $5/2^{+}$ low-lying states. The
two low-lying $5/2^{+}$ states contribute for almost $50\%$ of the total average number of nucleons in the decorrelated
region, in spite of their higher excitation energy than the two $1/2^{+}$ states (see Tab.~\ref{tab:Cr_spect}) and of the centrifugal confinement associated with their higher angular momentum. These hindrance effects are compensated by larger spectroscopic factors and by the intrinsic $6$-fold degeneracy of $5/2^{+}$ states. The significant contribution of $J=5/2$ states could not be expected from the usual qualitative picture that relates the emergence of a halo to $J=1/2$, or at most $J=3/2$, states.

The analysis method applied to neutron-rich Cr isotopes demonstrates unambiguously that a halo is predicted for
the last three bound isotopes. The halo region contains a small
fraction of neutrons that impact significantly the extension of the nucleus. It is generated by an admixture of four
 $J=1/2$ and $J=5/2$ states, which provides the picture of a rather \textit{collective} halo building up
at the neutron drip-line of Cr isotopes.

\subsection{Sn isotopes}
\label{sec:res_sn}

\begin{figure}[hptb]
\includegraphics[keepaspectratio,angle = -90, width = 0.5\columnwidth]{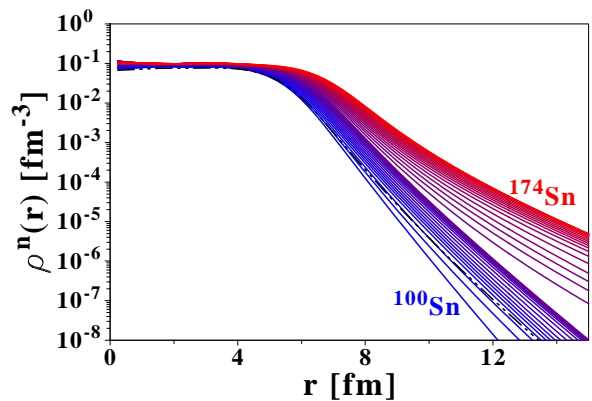}
\caption{(Color Online) Same as Fig.~\ref{fig:Cr_alldens} for Sn isotopes. The "separation" between the two groups of neutron densities occurs for \mbox{$N=82$}. Proton density of \mbox{$^{100}$Sn} is given as a reference in dashed-dotted line. Taken from Ref.~\cite{Rotival:2007hp}.}
 \label{fig:Sn_alldens}
\end{figure}

We now turn to drip-line tin isotopes as paradigmatic non-halo nuclei. At first glance, the long tail of the neutron density does exhibit a qualitative change for \mbox{$N>82$}, as seen in Fig.~\ref{fig:Sn_alldens}. However, the
transition is smoother than for chromium isotopes (Fig.~\ref{fig:Cr_alldens}). This rather smooth evolution is partly due the
increase of collectivity associated with the higher mass. There exist also specific nuclear-structure features that
explain the absence of halo in drip-line Sn isotopes.

\begin{figure}[hptb]
\includegraphics[keepaspectratio,angle = -90, width = 0.5\columnwidth]{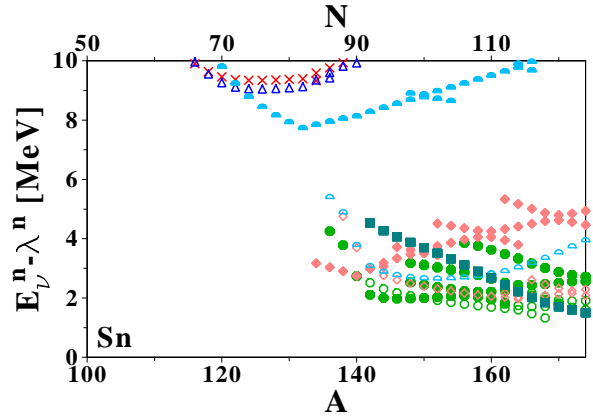}
\caption{ \label{fig:Sn_hfb_spectrum} (Color Online) Same as Fig.~\ref{fig:Cr_hfb_spectrum} for one-neutron separation energies of
Sn isotopes. Taken from Ref.~\cite{Rotival:2007hp}.}
\end{figure}

Table~\ref{tab:Sn_spect} details the one-neutron separation energy spectrum \mbox{$|E^{-}_\nu|$} between the $J^\pi=0^+$ ground state of the drip-line nucleus \mbox{$^{174}$Sn} and states in \mbox{$^{173}$Sn}. The energy scales at play are not compliant with the emergence of a halo, as is visually confirmed in Fig.~\ref{fig:Sn_hfb_spectrum}. The ground state whose separation energy is \mbox{$S_{n}=E\approx1.5$}~MeV is bunched with 10 low-lying states over a energy spread \mbox{$\Delta E\approx3.4$}~MeV. This group is separated from higher excitations
by \mbox{$E'\approx5.6$}~MeV. Thus, the energy spread of the low-lying states \mbox{$\Delta E$} is too
large to favor the formation of a halo. Additionally, according to the phenomenological criterion extracted from light halo nuclei,
the one-neutron separation energy of \mbox{$^{174}$Sn} should have been of the order of \mbox{$2$~MeV$/A^{2/3}\approx64$~keV}
for a halo to have a chance to emerge.

\begin{table}
\begin{tabular}{rll}
\hline 
&&\\
& $J^{\pi}$ & $|E^{-}_\nu|$~[MeV] \\
&&\\
\hline
&&\\
&   & $>15$ \\
& $5/2^{+}$ & 14.2 \\
& $3/2^{+}$ & 12.0 \\
& $1/2^{+}$ & 12.0 \\
& $11/2^{-}$ & 10.6 \\
\multirow{2}{*}{$E'\left\updownarrow
\vphantom{\begin{array}{l}a\\a\end{array}}\right.$}  &\\
&& \\
&& \\
\multirow{11}{*}{$\Delta E\left\{\vphantom{\begin{array}{l}a\\a\\a\\a\\a\\a\\a\\a\\a\\a\\a\end{array}}\right.$}
&$7/2^{-}$ & 4.9 \\
&$7/2^{-}$ & 4.5 \\
&$9/2^{-}$ & 3.9  \\
&$3/2^{-}$ & 2.7  \\
&$1/2^{-}$ & 2.6  \\
&$3/2^{-}$ & 2.5  \\
&$5/2^{-}$ & 2.3  \\
&$5/2^{-}$ & 2.1  \\
&$1/2^{-}$ & 1.9  \\
&$1/2^{-}$ & 1.6  \\
&$13/2^{+}$ & 1.5  \\
$E\updownarrow$& & \\
 \hline
\end{tabular}
\caption{\label{tab:Sn_spect} (Color Online) One-neutron separation energies \mbox{$|E^{-}_\nu|$} from the ground state of $^{174}$Sn to final states of $^{173}$Sn with spectroscopic factors greater than $10^{-2}$.}
\end{table}

\begin{figure}[hptb]
\includegraphics[keepaspectratio,angle = -90, width = 0.5\columnwidth]{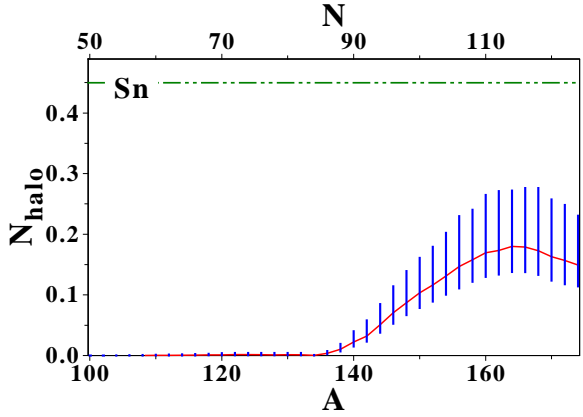}
\caption{(Color Online) Average number of nucleons in the spatially decorrelated region for Sn isotopes. For comparison,
\mbox{$N_{\mathrm{halo}}(^{80}$Cr$)$} is shown as a horizontal dashed-dotted line. Taken from Ref.~\cite{Rotival:2007hp}.} \label{fig:Sn_Nhalo}
\end{figure}

The $N_{\mathrm{halo}}$ parameter is displayed in Fig.~\ref{fig:Sn_Nhalo}. The maximum value of $N_{\mathrm{halo}}$, around $0.18$,
is very small compared to the total number of nucleons. The absolute numbers are also smaller than the ones
obtained in (lighter) drip-line Cr isotopes. The value of $N_{\mathrm{halo}}$ found here is of the same order of magnitude as those encountered for a non-halo $p$-wave nucleus such as $^{13}$N, where around $0.12$ neutron out of six reside in average in the classically forbidden region~\cite{nunespriv}. An interesting feature is the decrease of $N_{\mathrm{halo}}$ for \mbox{$N>166$}. This relates to the overlap function dominating the density at long distances being related to a high angular momentum $J^{\pi}=13/2^+$ ground state right in the last bound isotopes (see Fig.~\ref{fig:Sn_hfb_spectrum}). The spatial extension of this lowest-lying overlap function experiences a strong hindrance from the centrifugal barrier. As neutrons are added beyond $N=166$, the spectroscopic factor of this highly degenerate ground state increases, thus re-enforcing the localization of the density distribution up to $^{174}$Sn. The occurrence of such high-spin ground states and their hindrance effect on the halo formation is a marked difference between mid-mass and light nuclei.

\begin{figure}[hptb]
\includegraphics[keepaspectratio,angle = -90, width = 0.5\columnwidth]{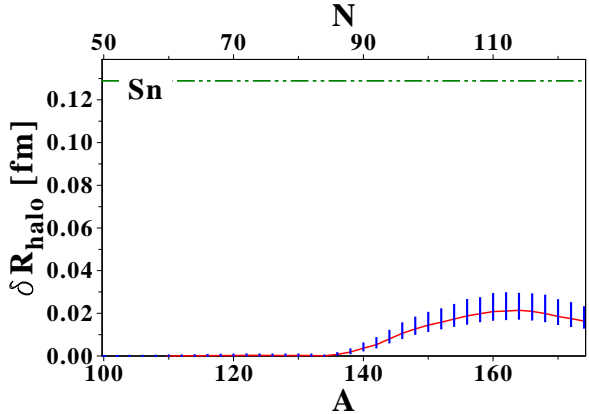}
\caption{(Color Online) Halo factor parameter $\delta{R}_{\mathrm{halo}}$ in the Sn isotopic chain. For comparison purposes, the maximum value of
$\delta R_{\mathrm{halo}}$ obtained for Cr isotopes is represented as a horizontal dashed-dotted line. Taken from Ref.~\cite{Rotival:2007hp}.}
\label{fig:Sn_deltaR}
\end{figure}

This picture is confirmed by displaying $\delta{R}_{\mathrm{halo}}$ in Fig.~\ref{fig:Sn_deltaR}. Indeed, the decorrelated
region is shown to have little influence on the nuclear extension, of the order of $0.02$~fm. The heavy mass of tin isotopes hinders the possibility of a sharp separation of core and tail contributions in the total density and thus of the formation of a halo, only allowing for a neutron skin to develop. 

\begin{figure}[hptb]
\includegraphics[keepaspectratio,angle = -90, width = 0.5\columnwidth]{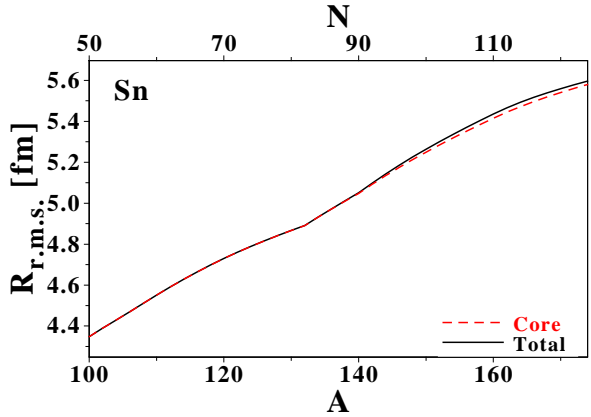}
\caption{(Color Online) Same as Fig.~\ref{fig:Cr_helmlike} for Sn isotopes. Taken from Ref.~\cite{Rotival:2007hp}.} \label{fig:Sn_helmlike}
\end{figure}

\section{Impact of pairing correlations}
\label{features}

The paradigmatic results presented in Sec.~\ref{appli} have been obtained with a particular EDF and it is of interest to probe the sensitivity of the predictions to the different ingredients of the method. In the present contribution, we limit the discussion to the impact of pairing correlations and of the specific parameterization used in the corresponding part of the nuclear EDF. For a discussion of the (significant\footnote{Based on acceptable variations of the parameters, up to 100 percent variations of $N_{\mathrm{halo}}$ and $\delta R_{\mathrm{halo}}$ can be obtained. At the level of our current knowledge of the parameterization of the nuclear EDF, these quantities thus appears to be very fined tuned.}) dependence of the predictions on the parametrization of the particle-hole part of the EDF driving the effective individual motion, see Ref.~\cite{Rotival:2007sy}.

\subsection{Pairing anti-halo effect?}
\label{sec:antihalo}

In the presence of pairing correlations, the asymptotic of the one-neutron density distribution takes a different form from the one it has in the EDF treatment based on an auxiliary Slater determinant~\cite{rotival07a}.Indeed, in first approximation, paired densities decrease faster than unpaired ones and pairing correlations induce an \textit{anti-halo effect} by localizing the one-body density~\cite{bennaceur99,bennaceur00,yamagami05}.

\begin{figure}[hptb]
\includegraphics[keepaspectratio, angle = -90, width = 0.5\columnwidth]{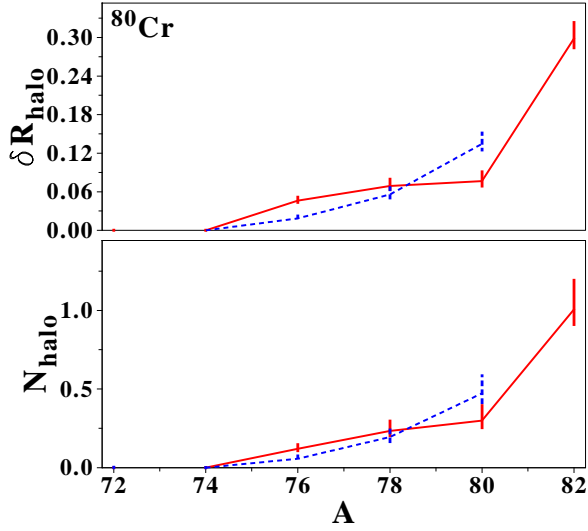}
\caption{ \label{fig:Cr_HF2} (Color Online) $N_{\mathrm{halo}}$ and $\delta R_{\mathrm{halo}}$ for Cr isotopes without (solid lines) and with (dashed lines) explicit treatment of pairing correlations. Taken from Ref.~\cite{Rotival:2007sy}.}
\end{figure}

To evaluate the quantitative impact of this effect, drip-line Cr isotopes have been calculated with and without
explicit treatment of pairing correlations. Firstly, \mbox{$^{82}$Cr} is predicted to be bound when pairing
correlations are excluded from the treatment, e.g. pairing correlations can change the position of the drip-line and modify in this way the number of halo candidates over the nuclear chart. Secondly, the values of $N_{\mathrm{halo}}$ and $\delta R_{\mathrm{halo}}$ are noticeably different in both cases, i.e. the neutron halo is significantly quenched in \mbox{$^{80}$Cr} when pairing is omitted whereas the situation is reversed in the
lighter isotopes, as seen in Fig.~\ref{fig:Cr_HF2}.

\begin{figure}[hptb]
\includegraphics[keepaspectratio, angle = -90, width = 0.5\columnwidth]{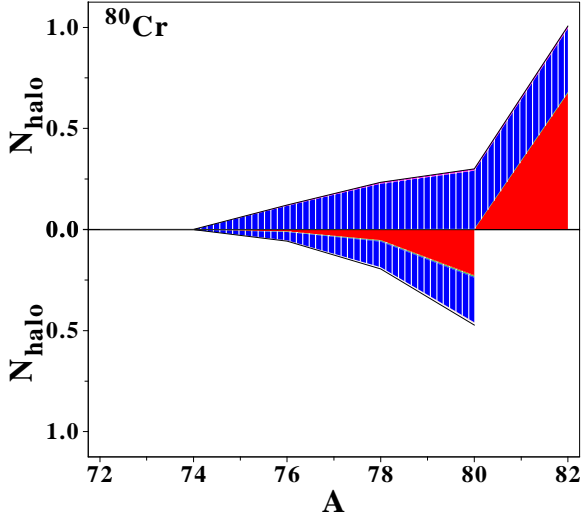}
\caption{ \label{fig:Cr_HF3} (Color Online) Added contributions to $N_{\mathrm{halo}}$ from the overlap functions associated with the two lowest $J^{\pi}=5/2^{+}$ (vertically-dashed areas) states and the two lowest $J^{\pi}=1/2^{+}$ (filled areas) states. Other contributions are found to be negligible. Top panel: without pairing correlations included. Bottom panel: with pairing correlations included. Taken from Ref.~\cite{Rotival:2007sy}.}
\end{figure}

Such results underline that pairing correlations affect halos in two opposite ways, i.e. they (i) inhibit the
formation of halos through the anti-halo effect while they may (ii) promote the formation of halos by enhancing spectroscopic factors of states having the lowest separation energies. For example, the anti-halo effect dominates in $^{76}$Cr and $^{78}$Cr whereas the pairing-induced increase of spectroscopic factors associated with the two very low-lying $J^{\pi}=1/2^{+}$ states (see Fig.~\ref{fig:Cr_hfb_spectrum}) makes the halo to be more pronounced in \mbox{$^{80}$Cr}. This is illustrated in Fig.~\ref{fig:Cr_HF3} where the added contributions to $N_{\mathrm{halo}}$ from the lowest $J^{\pi}=5/2^{+}$ and $J^{\pi}=1/2^{+}$ states are compared for calculations including or excluding pairing correlations. 


There is no simple answer as to whether pairing correlations enhance or hinder the formation of halos. The net result depends on structure details of the particular nucleus of interest~\cite{grasso01a}. See Ref.~\cite{Rotival:2007sy} for a more detailed discussion.

\subsection{Importance of low densities?}
\label{sec:pair_loc}

The local (neutron) pairing functional deriving from DDDI takes the form
\begin{equation}
{\cal E}^{{\rm pairing}} = \frac{V_0}{4} \, \left[1-\eta\left(\frac{\rho (\vec{r}\,)}{\rho_{\mathrm{sat}}} \right)^{\alpha}
\right] 
  |\tilde{\rho} (\vec{r}\,)|^{2}\, \, , \label{eq:Vpp}
\end{equation}
where $\tilde{\rho} (\vec{r}\,)$ denotes the local neutron pairing density. In addition to the strength $V_0$, two parameters $\eta$ and $\alpha$ control the spatial dependence of the effective coupling constant. With $\rho_{\mathrm{sat}}$ designating the saturation density of infinite nuclear matter, a zero value of $\eta$ corresponds to a pairing strength that is uniform over the nuclear volume (``volume
pairing'') while \mbox{$\eta=1$} corresponds to a pairing strength that is stronger in the vicinity of the nuclear
surface (``surface pairing''). A value $\eta=1/2$ corresponds to the intermediate situation (``mixed-type
pairing'') used so far in this paper. The parameter $\alpha$ was set to 1 so far as it is usually done. Values
\mbox{$\alpha<1$} correspond to stronger pairing correlations at low density, i.e. in the tail of the density distribution. 

All in all, $\eta$ and $\alpha$ strongly affect the spatial localization of the pairing field, and thus the gaps of weakly-bound orbitals lying at the nuclear surface. It is of interest to vary these empirical parameters to quantify how much the characteristics of the pairing functional impact halo systems. Properties of the last bound Cr isotopes are thus evaluated for different pairing functionals: (i)
\mbox{$\alpha=1$} and \mbox{$\eta\in[0,1]$}, along with (ii) \mbox{$\eta=1$} and \mbox{$\alpha\in[1,0.1]$}. The strength $V_0$ is systematically chosen so that the neutron spectral gap \mbox{$\langle\Delta_\kappa^n\rangle$}~\cite{doba96} equals $1.250$~MeV in \mbox{$^{120}$Sn}. Such a value of $V_0$ provides reasonable pairing gaps in Ca, Sn and Pb regions.

\begin{figure}[hptb]
\includegraphics[keepaspectratio, angle = -90, width = 0.5\columnwidth]%
{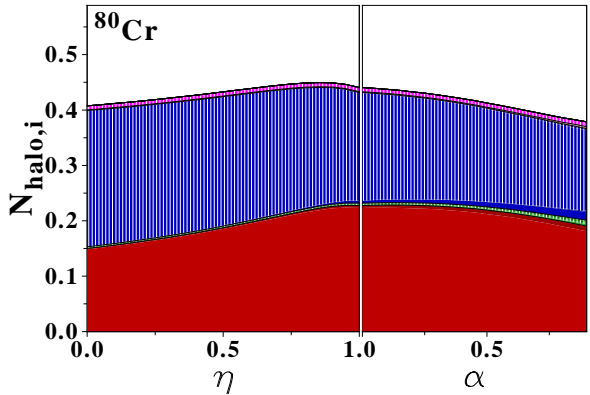}
\caption{(Color Online) Contributions to $N_{\mathrm{halo}}$ from the overlap functions associated with the two lowest $J^{\pi}=5/2^{+}$ (vertically-dashed areas) states and the two lowest $J^{\pi}=1/2^{+}$ (filled areas) states when varying $\eta$ and $\alpha$ (see text).  Taken from Ref.~\cite{Rotival:2007sy}. \label{fig:Cr80_RDFTX_decomp}}
\end{figure}

Within theoretical error bars, $N_{\mathrm{halo}}$ and $\delta R_{\mathrm{halo}}$ are essentially independent of $\eta$ and $\alpha$, although the anti-halo effect becomes more effective as $\alpha$ decreases. $N_{\mathrm{halo}}$ and $\delta R_{\mathrm{halo}}$ are maximal for the standard surface pairing functional \mbox{$\eta=\alpha=1$}. This translates into  the composition of the neutron halo in \mbox{$^{80}$Cr} displayed in Fig.~\ref{fig:Cr80_RDFTX_decomp}. 

In conclusion, the impact of the low density characteristics of the pairing functional on halo properties is found to be small, as long as the adequate renormalization scheme is used\footnote{See Ref.~\cite{Rotival:2007sy} for a thorough discussion of this point.}. Consequently, experimental constraints on pairing localization and the effective pairing strength based solely on halo properties are unlikely.

\section{Large scale predictions for semi-magic nuclei}
\label{predictions}

We are now in position to present large scale predictions of halo structures in mid-mass nuclei based on the one EDF parametrization used throughout the present contribution. We restrict ourselves to even-even semi-magic nuclei, which translates into about $500$ nuclei~\cite{hilaire07,doba04b}.

\begin{figure*}
\centering
\includegraphics[keepaspectratio,angle = -90, width = 0.6 \textwidth]%
{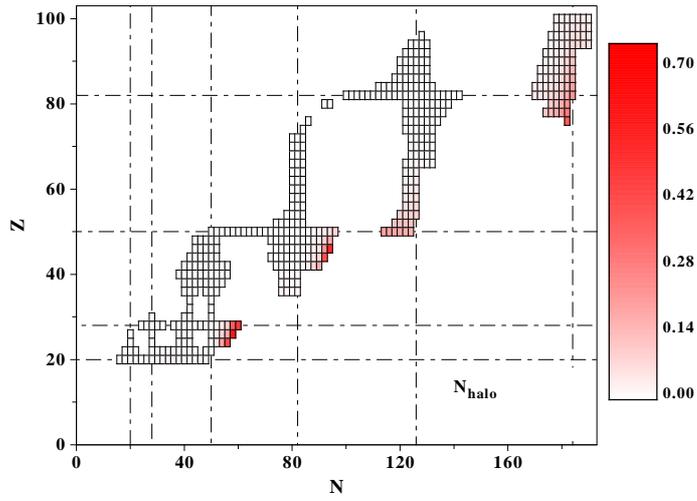} \caption{(Color Online) $N_{\mathrm{halo}}$ parameter predicted for about $500$ even-even semi-magic nuclei.  Taken from Ref.~\cite{Rotival:2007sy}.\label{fig:syst_nhalo}}
\end{figure*}

\begin{figure*}
\centering
\includegraphics[keepaspectratio,angle = -90, width = 0.6\textwidth]%
{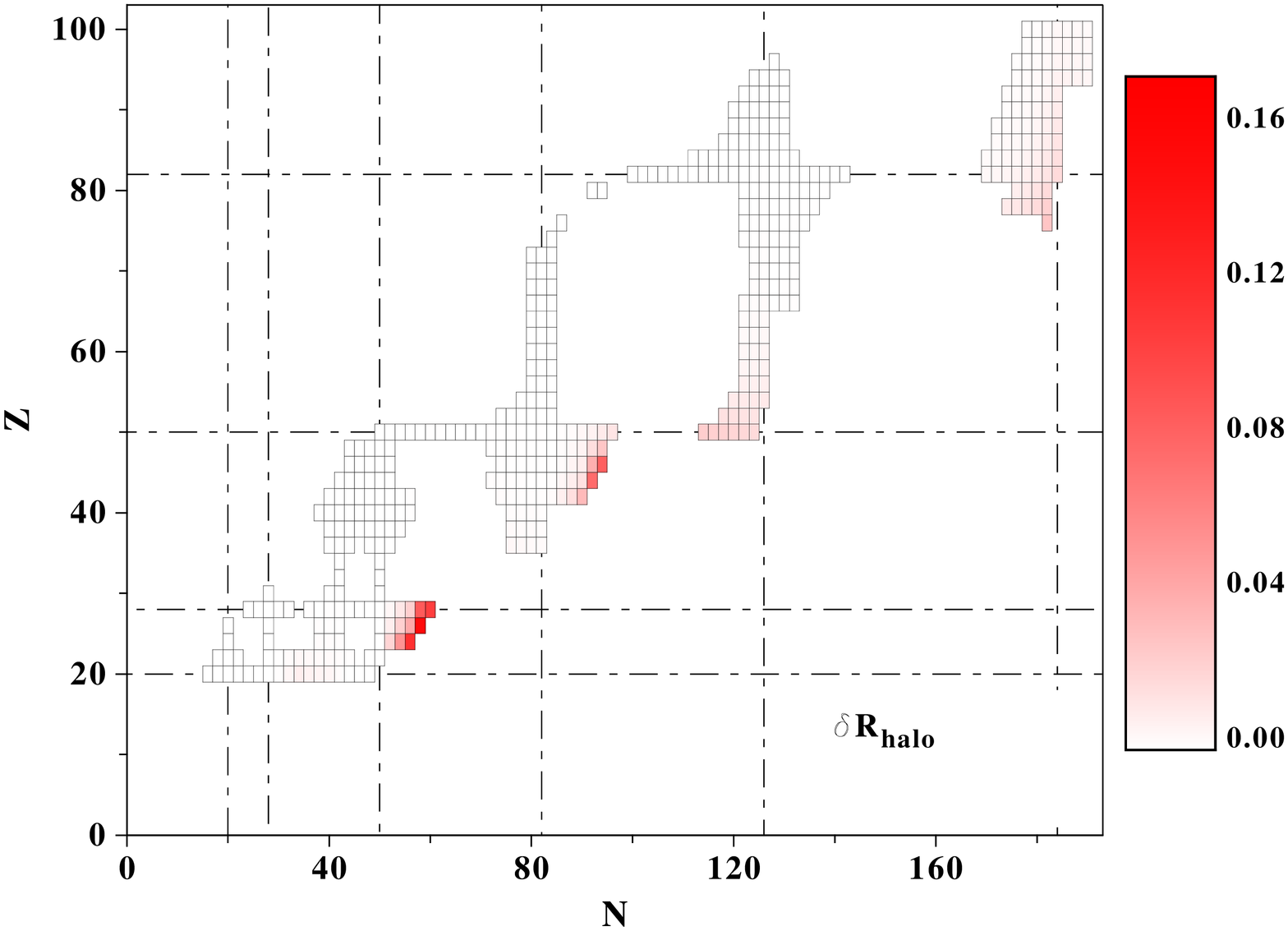} \caption{(Color Online) $\delta R_{\mathrm{halo}}$ parameter predicted for about $500$ even-even semi-magic nuclei. Taken from Ref.~\cite{Rotival:2007sy}.} \label{fig:syst_drhalo}
\end{figure*}

\begin{figure*}
\centering
\includegraphics[keepaspectratio,angle = -90, width = 0.5\textwidth]%
{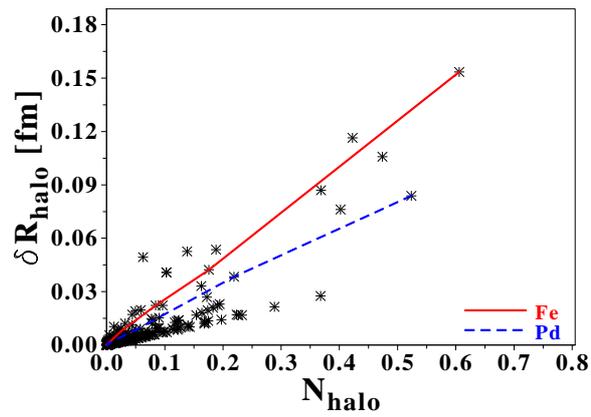} \caption{(Color Online) Correlation between $N_{\mathrm{halo}}$ and $\delta R_{\mathrm{halo}}$ for
about $500$ even-even semi-magic nuclei. The evolution of $\delta R_{\mathrm{halo}}$ as a function
of $N_{\mathrm{halo}}$ for Fe and Pd isotopes are highlighted. Taken from Ref.~\cite{Rotival:2007sy}.} \label{fig:phase}
\end{figure*}

Results for $N_{\mathrm{halo}}$ are shown in Fig.~\ref{fig:syst_nhalo}. We observe that (i) several isotopic chains are
predicted to display neutrons halos, (ii) halos are predicted to exist only at the very limit of neutron stability, i.e. typically less than four neutrons away from the drip line, (iii) the maximum value of $N_{\mathrm{halo}}$ is about $\sim0.7$, (iv) very few heavy elements in the (Pt, Hg, Tl...)
region are found to have a non-zero halo parameter $N_{\mathrm{halo}}$, (v) on the large scale, the halo phenomenon is very
rare and localized, (vi) looking at the best cases between $Z=20$ to $Z=100$, the absolute and relative
values of $N_{\mathrm{halo}}$ decrease with nuclear mass.

Results for $\delta R_{\mathrm{halo}}$ are presented in Fig.~\ref{fig:syst_drhalo} and confirm the above analysis on
$N_{\mathrm{halo}}$. In particular, it is seen that the fraction of decorrelated nucleon has almost no influence on the
nuclear extension of massive nuclei. Only two very localized regions where the predicted halo significantly
affects the neutron r.m.s. radius are found, e.g. for (i) Cr, Fe and Ni nuclei on the one hand, and (ii) Pd and Ru isotopes on the other hand. The drip-line isotopes of these elements are predicted as the best halo candidates for the presently used EDF parameterization. These
even-even nuclei have in common the presence of low-lying $J=1/2$ and $J=3/2$ states in the odd-even isotopes with one less neutron. Although states with larger angular momentum contribute to the nuclear halo in some cases as discussed above, the presence of those low-lying $J=1/2$ and/or $J=3/2$ states remains mandatory for a significant halo to emerge. That being said, no halo made \mbox{purely} out of $J=1/2$ and/or $J=3/2$ states has been found, illustrating the probable greater collective character of mid-mass halos as compared to well-studied light halo systems.

The complementarity between the two criteria $N_{\mathrm{halo}}$ and $\delta R_{\mathrm{halo}}$ appears more clearly through the
large scale analysis. The plot presented in Fig.~\ref{fig:phase} shows that the two observables are correlated
within a given isotopic chains such that the information carried by both quantities is somewhat redundant in this case. On a larger scale however, the
correlation pattern changes as the proton number increases (between Cr and Pd isotopes for instance), i.e.
$\delta R_{\mathrm{halo}}$ increases much less with $N_{\mathrm{halo}}$ as the mass increases.

\section{Conclusions}
\label{conclusions}

Looking into the future, it is interesting to question the possible occurrence of halos beyond well-studied p-shell or light sd-shell nuclei. As the neutron drip-line is only known experimentally up to oxygen, one is currently limited to speculating theoretically at this point in time, awaiting for experimental answers to arise with the upcoming generation of radioactive ion beam facilities. 

In view of doing so, the present contribution first reviewed the structure analysis method recently proposed in Ref.~\cite{Rotival:2007hp} and applied systematically in Ref.~\cite{Rotival:2007sy}. The method is solely based on the one-nucleon density distribution and applies to all systems from non halos to paradigmatic halos without any a priori consideration. The method eventually materialize into quantitative measures of the average number of nucleons participating in the halo and of the impact the halo region has on the nuclear spatial extension. The emergence of a neutron halo in even-even mid-mass nuclei is shown to be related to three typical energy scales characterizing the spectrum of the isotope with one less neutron.

In a second part, the analysis method was validated via results obtained from single-reference energy density functional calculations of chromium and tin drip-line isotopes, the former (latter) playing the role of paradigmatic (non) halo candidates. Halos at the drip line of chromium isotopes were shown to be generated by an admixture of overlap functions associated with four low-lying $J=1/2$ and $J=5/2$ states in the neighboring isotope with one less neutron, thus providing the picture of rather \textit{collective} halos. The halo region was shown to contain a small fraction of neutrons on average having a significant impact on the nuclear extension.

In a third part, the sensitivity of the predictions to the nuclear energy density functional employed was illustrated before providing large-scale predictions of halo properties on over $500$ even-even semi-magic nuclei. At the level of our current knowledge of the nuclear energy density functional parameterization, the quantities used to characterize the halo appear to be very fined tuned, i.e. they vary greatly as a result of modest modifications of (some of) the characteristics of the energy density functional parameterization. Of course, it would be of interest to apply the analysis method to results generated in mid-mass nuclei via other many-body techniques, e.g. ab initio many-body methods applicable to open-shell mid-mass nuclei such as multi-reference in-medium similarity renormalization group~\cite{Hergert:2013uja,Hergert:2014vn}, self-consistent Gorkov Green's function~\cite{soma11a,Soma:2013xha} or Bogoliubov coupled cluster~\cite{Signoracci:2014dia} theories.

On the large scale, the halo phenomenon was shown to be very rare and localized, drip line nuclei of $Z\approx 24-28$ and $Z\approx44-46$ elements being predicted as the best halo candidates for the  EDF parameterization presently employed. While states with larger angular momentum contribute to the nuclear halo in some cases, the presence of those low-lying $J=1/2$ and/or $J=3/2$ states was shown to remain mandatory for a significant halo to emerge. Eventually, drip-line isotopes of $Z\approx 44-46$ elements appear as the heaviest nuclei beyond which the potential halo has no significant influence on its spatial extension. However, not only this conclusion is based on simulations limited to even-even semi-magic nuclei but the impact such a potential halo may still have on other structure observable and reaction processes remains to be investigated.

\begin{acknowledgements}
The author would like to thank V. Rotival and K. Bennaceur deeply for collaborating on the matter that are reviewed in the present contribution.
\end{acknowledgements}

\bibliographystyle{spbasic}
\bibliography{mid_mass_neutron_halo}   

\end{document}